\documentstyle[11pt,newpasp,twoside,epsf]{article}

\newcommand{\MgIIdblt}{{\rm Mg}\kern 0.1em{\sc ii}~$\lambda\lambda 2796, 2803$}

\def\edcomment#1{\iffalse\marginpar{\raggedright\sl#1\/}\else\relax\fi}
\reversemarginpar

\begin{document}
\title{Phase Structure of Weak {\hbox{{\rm Mg}\kern 0.1em{\sc ii}}} Absorbers: Star Forming Pockets Outside of Galaxies}
\author{Jane C. Charlton, Christopher W. Churchill, Jie Ding, Stephanie Zonak, Nicholas Bond}
\affil{Department of Astronomy and Astrophysics, Penn State, University Park, PA 16802}
\author{Jane R. Rigby}
\affil{Astronomy Department, University of Colorado, Campus Box 391, Boulder, CO 80309}

\begin{abstract}
A new and mysterious class of object has been revealed by the
detection of numerous weak {\hbox{{\rm Mg}\kern 0.1em{\sc ii}}}
doublets in quasar absorption line spectra.  The properties of these
objects will be reviewed.  They are not in close proximity to luminous
galaxies, yet they have metallicities close to the solar value; they
are likely to be self-enriched.  A significant fraction of the weak
{\hbox{{\rm Mg}\kern 0.1em{\sc ii}}} absorbers are constrained to be
less than 10 parsecs in size, yet they present a large cross section
for absorption, indicating that there are a million times more of them
than there are luminous galaxies.  They could be remnants of
Population III star clusters or tracers of supernova remnants in a
population of ``failed dwarf galaxies" expected in cold dark matter
structure formation scenarios.
\end{abstract}

\section{Introduction}
Quasar absorption line systems were once thought to be distinct
classes of objects, such as the {\hbox{{\rm Ly}\kern 0.1em$\alpha$}}
forest, {\hbox{{\rm C}\kern 0.1em{\sc iv}}} systems, or {\hbox{{\rm
Mg}\kern 0.1em{\sc ii}}} systems.  Now it is recognized that there is
considerable overlap between these classes.  The remaining challenge
is to achieve complete understanding of the relationship between the
absorption line systems and the gaseous galaxies and intergalactic
structures in which they arise.

A clear association has been established between strong {\hbox{{\rm
Mg}\kern 0.1em{\sc ii}}} absorption lines and relatively luminous
galaxies, $>0.1 L_*$ (eg., Bergeron \& Boiss\'e 1991; Steidel,
Dickinson, \& Persson 1994; Steidel 1995).  However, the gaseous
structures giving rise to weak {\hbox{{\rm Mg}\kern 0.1em{\sc ii}}}
absorbers, those with rest frame equivalent widths, $W_r(2796)$, less
than $0.3$~{\AA} have not yet been identified.  In this contribution,
we use other chemical transitions such as {\hbox{{\rm Fe}\kern
0.1em{\sc ii}}}, {\hbox{{\rm C}\kern 0.1em{\sc iv}}}, {\hbox{{\rm
Si}\kern 0.1em{\sc iv}}} and {\hbox{{\rm Ly}\kern 0.1em$\alpha$}} to
determine if multiple ionization phases of gas are present and to
constrain their physical properties.  We then speculate about the
origin of the weak {\hbox{{\rm Mg}\kern 0.1em{\sc ii}}} absorbers.

\section{Statistics}
In Figure 1, we show examples of three single--cloud weak {\hbox{{\rm
Mg}\kern 0.1em{\sc ii}}} absorbers and one multiple cloud (bottom
right panel) weak {\hbox{{\rm Mg}\kern 0.1em{\sc ii}}} absorber.
\begin{figure}[th]
\plotfiddle{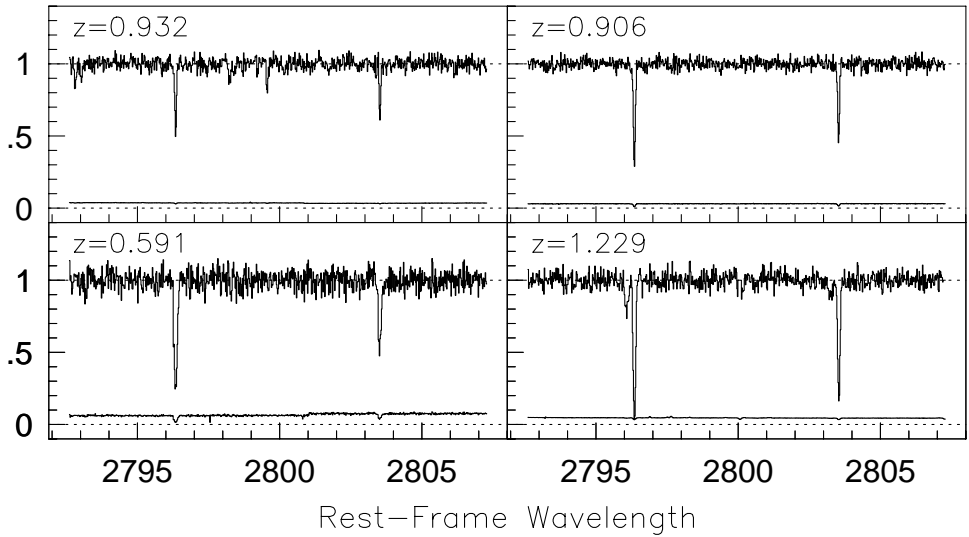}{2.7in}{0}{80.}{80.}{-255}{-380}
\caption{Representative ``weak" {\hbox{{\rm Mg}\kern 0.1em{\sc ii}}}
absorbers as seen in HIRES/Keck spectra (Churchill et al. 1999, ApJS,
120, 51).  The {{\rm Mg}\kern 0.1em{\sc ii}~$\lambda\lambda 2796,
2803$} doublet is displayed in the rest frame.  The sample includes
three single and one double (bottom right panel) component weak
absorbers.}
\end{figure}
At $0.5 < z < 1.0$, weak {\hbox{{\rm Mg}\kern 0.1em{\sc ii}}}
absorbers are more numerous than strong absorbers.  The equivalent
width distribution of {\hbox{{\rm Mg}\kern 0.1em{\sc ii}}} absorbers
is shown in Figure 2.
\begin{figure}[th]
\plotfiddle{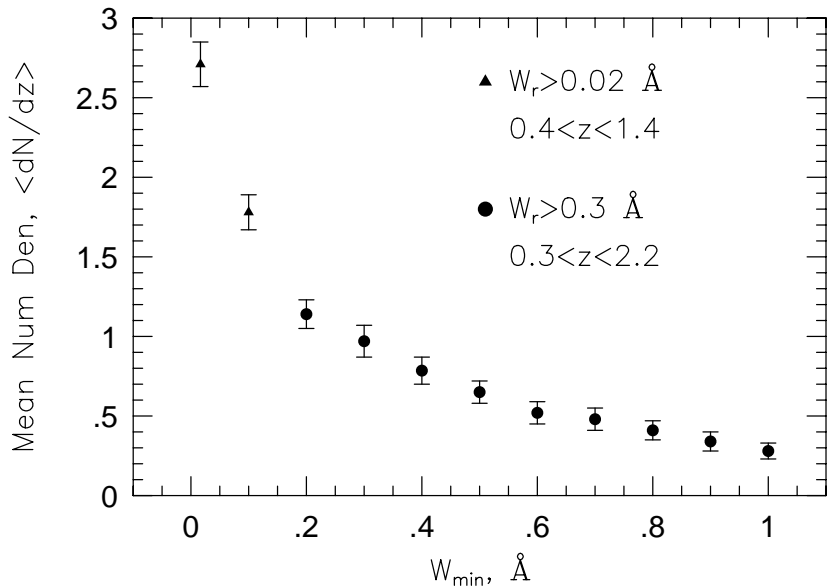}{3.00in}{0}{95.}{95.}{-210}{-300}
\caption{The number density of
absorbers with equivalent widths, $W$, greater than a well-defined
minimum, $W_{min}$, measured over finite redshift bins.  The data for
weak {\hbox{{\rm Mg}\kern 0.1em{\sc ii}}} absorbers are taken from
Churchill et al. (1999), and those for strong {\hbox{{\rm Mg}\kern
0.1em{\sc ii}}} absorbers from Steidel and Sargent (1992).}
\end{figure}
It rises rapidly toward weaker lines, and for $W_r(2796) < 0.3$~{\AA},
there are $dN/dz = 1.74\pm0.10$ systems per unit redshift.  However,
some of these systems have multiple clouds and are kinematically
similar to stronger {\hbox{{\rm Mg}\kern 0.1em{\sc ii}}} systems.
Considering only the single--cloud weak {\hbox{{\rm Mg}\kern 0.1em{\sc
ii}}} absorbers, we find $dN/dz = 1.16\pm0.07$.

\section{Comparing {\hbox{{\rm Mg}\kern 0.1em{\sc ii}}}, {\hbox{{\rm Fe}\kern 0.1em{\sc ii}}}, {\hbox{{\rm C}\kern 0.1em{\sc iv}}}, and {\hbox{{\rm H}\kern 0.1em{\sc i}}} Absorption}
Clues to the nature of single--cloud weak {\hbox{{\rm Mg}\kern
0.1em{\sc ii}}} absorbers are derived from a comparison of other
transitions detected at the same velocity.  Figure 3 shows, for
$15$ systems, the regions of the observed spectrum in which the key
transitions, {\hbox{{\rm Fe}\kern 0.1em{\sc ii}}}, {\hbox{{\rm C}\kern
0.1em{\sc iv}}}, and {\hbox{{\rm Ly}\kern 0.1em$\alpha$}}, as well as
the Lyman limit, would appear given the presence of weak {\hbox{{\rm
Mg}\kern 0.1em{\sc ii}}}.
\begin{figure}[pth]
\plotfiddle{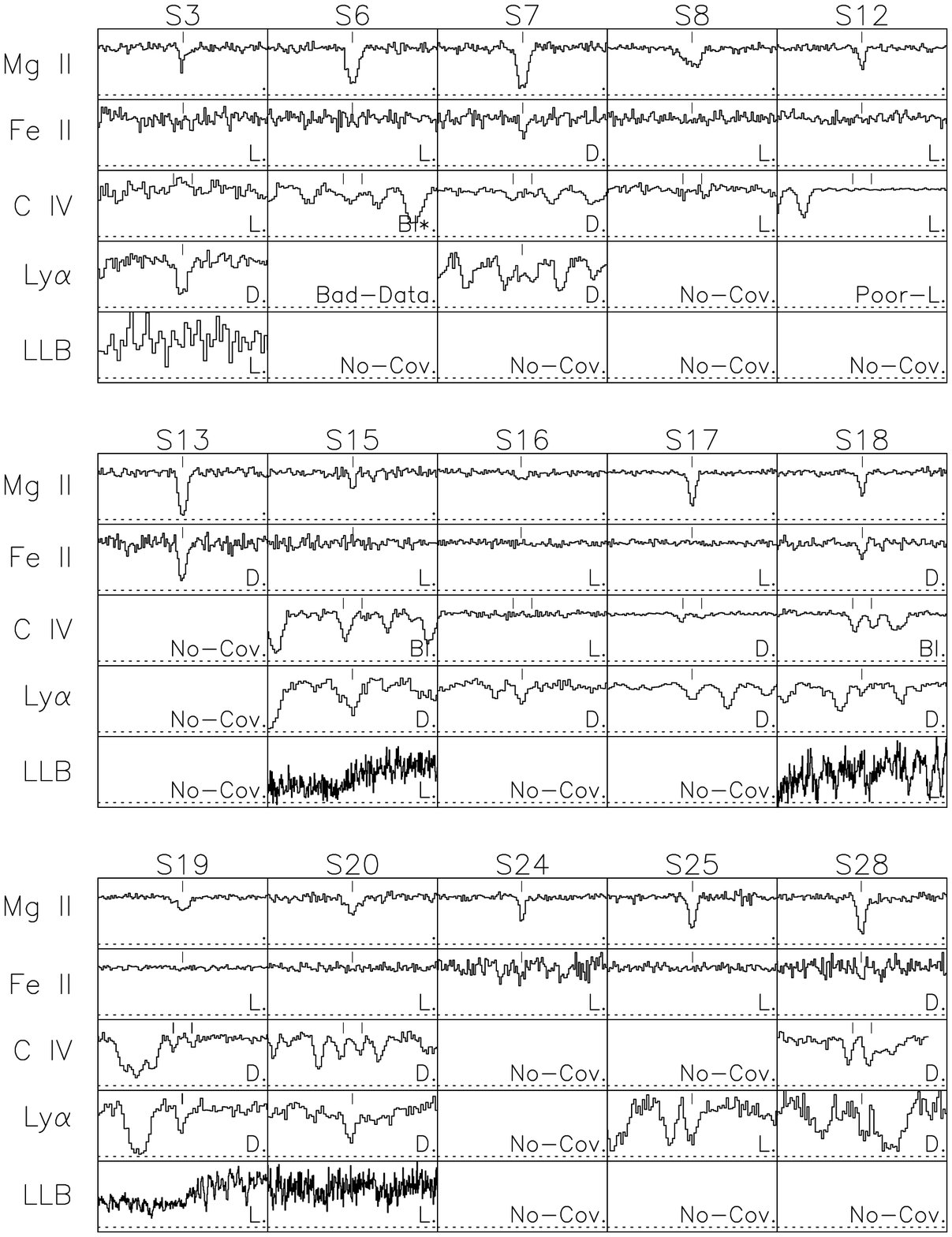}{6.0in}{0}{70.}{70.}{-225}{-60}
\caption{The ``data matrix'' for the single--cloud weak {\hbox{{\rm Mg}\kern 0.1em{\sc ii}}} systems.  For
each absorber, the {\hbox{{\rm Mg}\kern 0.1em{\sc ii}}} $\lambda 2796$
transition is shown in the top sub--panel.  In the lower sub--panels, we
present the spectral regions where the {\hbox{{\rm Fe}\kern 0.1em{\sc
ii}}} $\lambda 2600$ (or $\lambda 2383$) transition, the {\hbox{{\rm
C}\kern 0.1em{\sc iv}}} doublet, the {\hbox{{\rm Ly}\kern
0.1em$\alpha$}} transition, and the Lyman limit break are expected.
Ticks above the spectra give the locations where features are
expected.  The full velocity window of the sub--panels with {\hbox{{\rm
Mg}\kern 0.1em{\sc ii}}} and {\hbox{{\rm Fe}\kern 0.1em{\sc ii}}} from
HIRES/Keck is $100$~{\hbox{km~s$^{-1}$}} and for the FOS/HST data is
$5000$~{\hbox{km~s$^{-1}$}}.  ``No--Cov'' indicates that the spectral
region was not observed, and ``Bad--Data'' indicates that
signal-to-noise ratio in the spectral region was too low for a useful
measurement.  ``D'' indicates a clean detection at the $3\sigma$ or
greater significance level.  ``L'' denotes no detection, but only an
upper limit on the equivalent width.  ``Bl'' indicates poor
constraints due to blending with other features.  Transitions not
plotted can be found in Churchill et al. (2000).}
\end{figure}
The {\hbox{{\rm Mg}\kern 0.1em{\sc ii}}} and {\hbox{{\rm Fe}\kern
0.1em{\sc ii}}} absorption profiles were obtained with a resolution of
$\sim 6$~{\hbox{km~s$^{-1}$}} with HIRES/Keck (see Churchill et al.
1999).  The {\hbox{{\rm C}\kern 0.1em{\sc iv}}}, {\hbox{{\rm Ly}\kern
0.1em$\alpha$}}, and Lyman series coverage, with resolution
$230$~{\hbox{km~s$^{-1}$}}, was obtained with the Faint Object
Spectrograph (FOS) aboard the {\it Hubble Space Telescope} (HST),
primarily as part of the HST QSO Absorption Lines Key Project (Bahcall
et al. 1993; Bahcall et al. 1996; Jannuzi et al. 1998).

In Figure 3, three of the $15$ systems, labeled S7, S13, and S18,
have detected {\hbox{{\rm Fe}\kern 0.1em{\sc ii}}} with a column
density comparable to {\hbox{{\rm Mg}\kern 0.1em{\sc ii}}}.  For these
iron--rich single--cloud weak {\hbox{{\rm Mg}\kern 0.1em{\sc ii}}}
systems, $dN/dz \simeq 0.18$.  Seven of the $15$ systems are
constrained to have multiphase conditions.  For five of those (S7,
S17, S19, S20, and S28), the {\hbox{{\rm C}\kern 0.1em{\sc iv}}} is
too strong to arise in the same phase with the {\hbox{{\rm Mg}\kern
0.1em{\sc ii}}}.  For three systems (S3, S15, and S20), the
{\hbox{{\rm H}\kern 0.1em{\sc i}}} cannot be produced in the same
phase with the {\hbox{{\rm Mg}\kern 0.1em{\sc ii}}}.  This is
discussed in detail by Rigby, Charlton, and Churchill (2001) and the
key points are summarized below.

\section{Physical Conditions for Single--Cloud Weak {\hbox{{\rm Mg}\kern 0.1em{\sc ii}}} Absorbers}

\subsection{From Photoionization Models}
The photoionization code Cloudy (Ferland 1996) was used to infer the
metallicity and ionization condition in weak {\hbox{{\rm Mg}\kern
0.1em{\sc ii}}} absorbers.  Cloudy calculated the
densities of the different ions for various chemical elements, layer
by layer, as incident photons traveled through a slab of material.  The
input parameters for Cloudy were the metallicity and the ionization
parameter (ratio of the number density of photons to the number
density of electrons).  We assumed the ionizing radiation spectrum
from the extragalactic background was
the modified quasar spectrum of Haardt and Madau (1996).  Since the
photon number density is set, the ionization parameter translates
directly to an electron number density.  Figure 4 summarizes the
dependence of ratios of the constraining transitions, $N({\hbox{{\rm
Fe}\kern 0.1em{\sc ii}}})/N({\hbox{{\rm Mg}\kern 0.1em{\sc ii}}})$ and
$N({\hbox{{\rm C}\kern 0.1em{\sc iv}}})/N({\hbox{{\rm Mg}\kern
0.1em{\sc ii}}})$, on ionization parameter, $\log U$.  In the
optically thin regime, this is virtually independent of metallicity
because all layers of the slab are exposed to the same radiation.
\begin{figure}[th]
\plotone{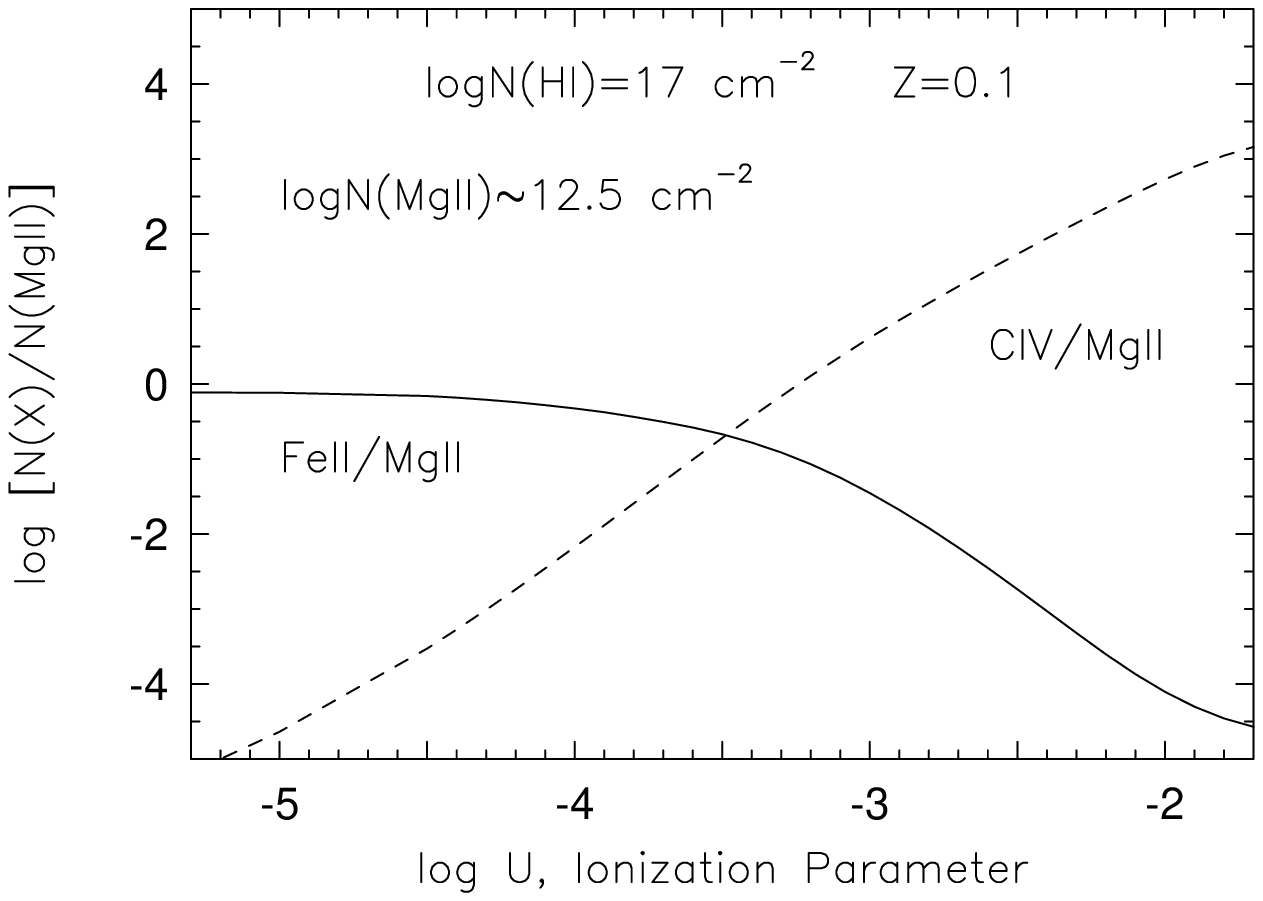}
\caption{The ratios $N({\hbox{{\rm Fe}\kern 0.1em{\sc ii}}})/N({\hbox{{\rm Mg}\kern 0.1em{\sc ii}}})$ and $N({\hbox{{\rm C}\kern 0.1em{\sc iv}}})/N({\hbox{{\rm Mg}\kern 0.1em{\sc ii}}})$ 
are uniquely determined functions of the ionization parameter over
three orders of magnitude in $\log U$.  Since ionization structure is
not important for weak {\hbox{{\rm Mg}\kern 0.1em{\sc ii}}} absorbers,
the ratios are independent of metallicity.  Note that at low values of
$\log U$, the {\hbox{{\rm Fe}\kern 0.1em{\sc ii}}}/{\hbox{{\rm
Mg}\kern 0.1em{\sc ii}}} ratio flattens and thus provides less
constraint.}
\end{figure}

The ratio $N({\hbox{{\rm Fe}\kern 0.1em{\sc ii}}})/N({\hbox{{\rm
Mg}\kern 0.1em{\sc ii}}})$ is quite sensitive to $\log U$ for $\log U
> -3.5$, but it flattens and does not provide a good constraint for
lower values of $\log U$.  For the three iron--rich systems shown in
Figure 3, $N({\hbox{{\rm Fe}\kern 0.1em{\sc ii}}})/N({\hbox{{\rm
Mg}\kern 0.1em{\sc ii}}})$ is close to one so that $\log U < -3.5$,
and therefore electron number density $\log n_e > 0.01 {\rm cm}^{-3}$
applies.  In these weak {\hbox{{\rm Mg}\kern 0.1em{\sc ii}}}
absorbers, the total column density is relatively small, and the high
density translates to a thickness of less than $10$~pc.

$N({\hbox{{\rm C}\kern 0.1em{\sc iv}}})/N({\hbox{{\rm Mg}\kern
0.1em{\sc ii}}})$ also provides a strong constraint on the ionization
parameter.  For some systems, the observed ratio is so large that
multiphase conditions are required (S7, S17, S19, S20, and S28 in
Figure 3).

For almost all single--cloud weak systems for which there is a
constraint, the metallicities are greater than $0.1$ solar, and in
many cases close to solar.  This constraint comes from a
{\hbox{{\rm Ly}\kern 0.1em$\alpha$}} line, which is weak compared to
the {\hbox{{\rm Mg}\kern 0.1em{\sc ii}}}, and/or from the absence of a
Lyman limit break.

\subsection{The Story of Three Single--Cloud Weak {\hbox{{\rm Mg}\kern 0.1em{\sc ii}}} Absorbers}
For three single--cloud weak {\hbox{{\rm Mg}\kern 0.1em{\sc ii}}}
absorbers along the line of sight toward the quasar PG~$1634+706$,
observations were obtained at high resolution ($R=30,000$) with the
Space Telescope Imaging Spectrograph (STIS) aboard HST.  The key
transitions for these three systems are compared in Figure 5.
\begin{figure}[pth]
\plotfiddle{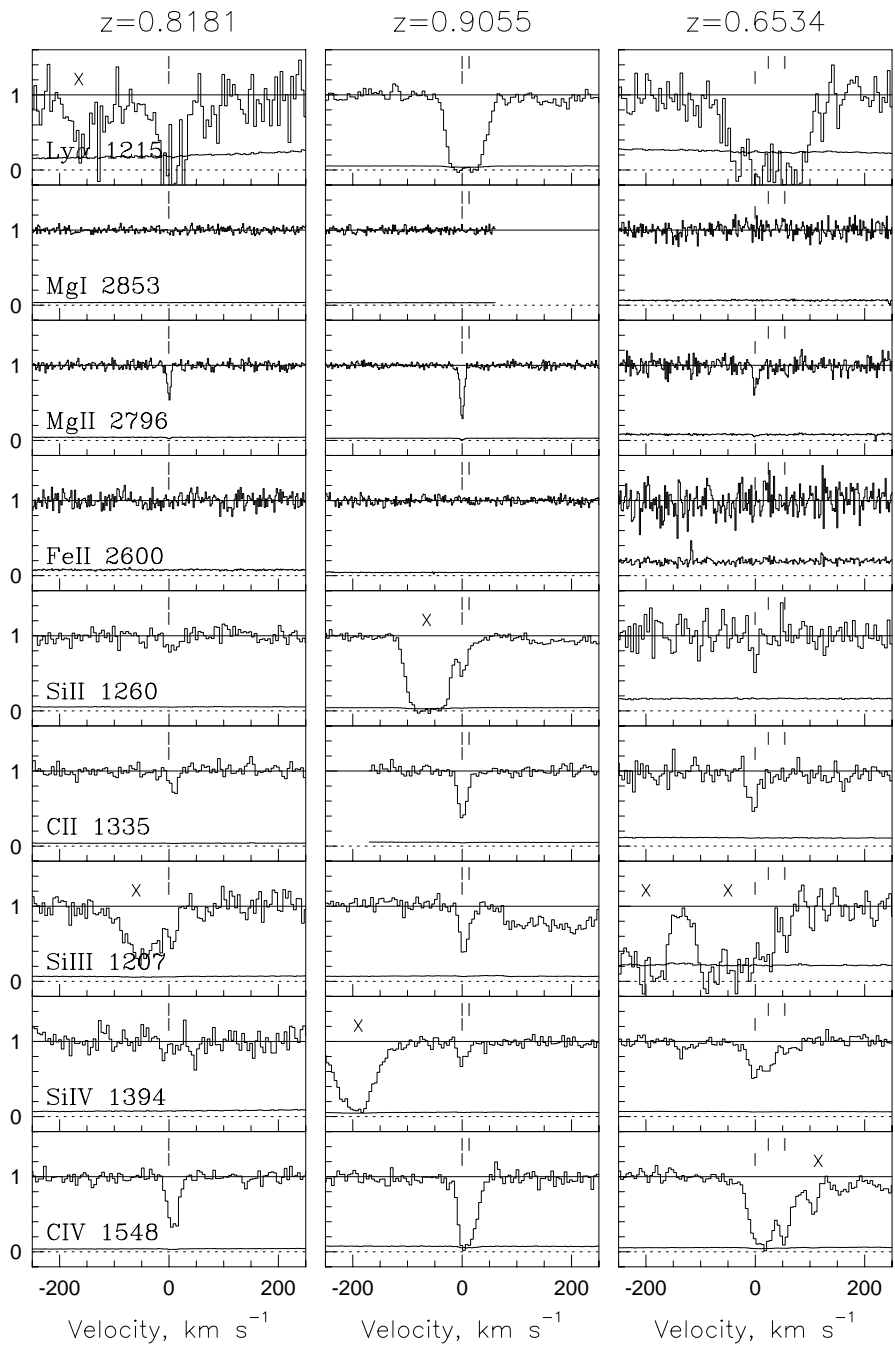}{7.0in}{0}{75.}{75.}{-240}{-40}
\caption{Comparison plot of key transitions for the three weak,
single--cloud {\hbox{{\rm Mg}\kern 0.1em{\sc ii}}} absorbers along the
PG~$1634+706$ line of sight.  Detected transitions and constraining
limits are presented in velocity space.  The data are at $R=45,000$
from HIRES/Keck for the {\hbox{{\rm Mg}\kern 0.1em{\sc ii}}},
{\hbox{{\rm Mg}\kern 0.1em{\sc i}}}, and {\hbox{{\rm Fe}\kern
0.1em{\sc ii}}} transitions.  All other transitions are taken from
STIS/HST spectra (combining observations by P.I.'s Scott Burles and
Buell Jannuzi), at resolution $R=30,000$.  The position of the lower
row of ticks, displayed above all of the transitions (at zero
velocity), was determined based upon a simultaneous Voigt profile fit
to the {\hbox{{\rm Mg}\kern 0.1em{\sc ii}}} doublet.  The upper row of
ticks show velocities of the additional components that were required
to fit {\hbox{{\rm C}\kern 0.1em{\sc iv}}}.}
\end{figure}
None of these three systems classify as iron--rich, because they
do not have detected {\hbox{{\rm Fe}\kern 0.1em{\sc ii}}}.
Methods of modeling these three systems, and details of the results,
are described in Charlton et al. (2001).  A brief summary follows.

In the $z=0.8182$ system, the {\hbox{{\rm C}\kern 0.1em{\sc iv}}}
absorption requires a second broader phase ($b \simeq
10$~{\hbox{km~s$^{-1}$}}) centered at the same velocity as the phase
that produces the weak {\hbox{{\rm Mg}\kern 0.1em{\sc ii}}} absorption
(with $b \simeq 2$~{\hbox{km~s$^{-1}$}}).  The relatively weak
{\hbox{{\rm Ly}\kern 0.1em$\alpha$}} constrains the system to have
solar metallicity.  The low ionization phase has a relatively high
density, $\sim 0.1$~{\hbox{cm$^{-3}$}}, and therefore a small size
(parsec scale).  The high ionization phase is hundreds of parsecs in
size, as constrained by {\hbox{{\rm Si}\kern 0.1em{\sc iv}}} and
{\hbox{{\rm N}\kern 0.1em{\sc v}}}, as well as {\hbox{{\rm C}\kern
0.1em{\sc iv}}}.  The simplest physical interpretation of these
results is a small dense region embedded in a larger more diffuse
region.  However, in a photoionized model the phases are not in
pressure balance because their temperatures are nearly equal.

In the $z=0.9056$ system, the {\hbox{{\rm C}\kern 0.1em{\sc iv}}} is
stronger than in the $z=0.8182$ system.  The metallicity is
constrained to be solar or higher.  It requires a separate higher
ionization phase centered on the {\hbox{{\rm Mg}\kern 0.1em{\sc ii}}},
but it also shows an asymmetric structure indicating that an
additional cloud $\sim 13$~{\hbox{km~s$^{-1}$}} to the red is
required.  For this system, the low ionization phase is of lower
density, $\sim 0.01$~{\hbox{cm$^{-3}$}}, than in the $z=0.8182$
system, and its size is $30$ to $100$~pc.  The high ionization phase
is somewhat larger.  Physically, the lower ionization phase could be
embedded within the higher ionization phase, or the high ionization
gas could be in the interior of a cold shell (fragmented so that we do
not see two cold components from opposite sides of the shell).  The
additional cloud could arise from gas physically separated from the
other material.

The $z=0.6564$ system has considerably stronger {\hbox{{\rm Ly}\kern
0.1em$\alpha$}} absorption than the other two systems, indicating a
lower metallicity of less than a tenth solar.  This appears to be
quite unusual since most single--cloud weak {\hbox{{\rm Mg}\kern
0.1em{\sc ii}}} absorbers have close to solar metallicity.  The
{\hbox{{\rm C}\kern 0.1em{\sc iv}}} in this system has complex
structure, with two offset clouds (at $24$ and
$54$~{\hbox{km~s$^{-1}$}}) that are more highly ionized and
do not give rise to {\hbox{{\rm Mg}\kern 0.1em{\sc ii}}} absorption.
The absorption profiles at the position of the {\hbox{{\rm Mg}\kern
0.1em{\sc ii}}} cloud are best fit with two
phases, because the {\hbox{{\rm C}\kern 0.1em{\sc iv}}} profiles is
broad relative to the {\hbox{{\rm Mg}\kern 0.1em{\sc ii}}}.

\subsection{Comparison to the {\hbox{{\rm Ly}\kern 0.1em$\alpha$}} Forest}
The single--cloud weak {\hbox{{\rm Mg}\kern 0.1em{\sc ii}}} absorbers
are mostly sub-Lyman limit systems, with $\log N({\hbox{{\rm H}\kern
0.1em{\sc i}}}) < 16.8$~{\hbox{cm$^{-2}$}.  This is shown
statistically in the left hand panel of Figure 6.
\begin{figure}[th]
\plotfiddle{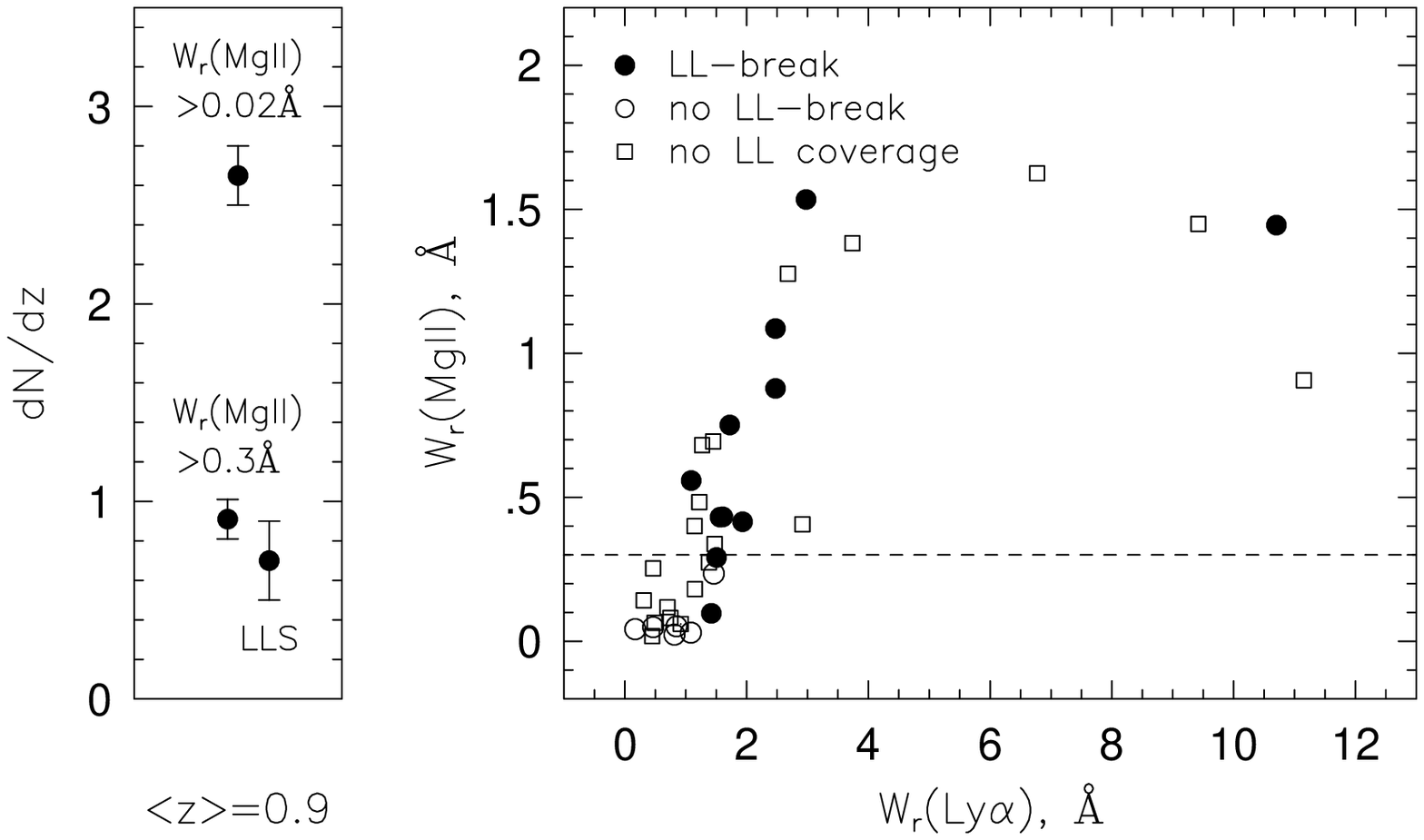}{3.0in}{0}{80.}{80.}{-260}{-340}
\caption{(left)- The redshift number density of Lyman
limit systems (LLSs), strong {\hbox{{\rm Mg}\kern 0.1em{\sc ii}}}
systems, and all {\hbox{{\rm Mg}\kern 0.1em{\sc ii}}} systems
including weak systems. Taken from Churchill et al. (1999). (right)-
The {\hbox{{\rm Mg}\kern 0.1em{\sc ii}}} -- {\hbox{{\rm Ly}\kern
0.1em$\alpha$}} equivalent width plane showing the optically thick
(solid circles) and optically thin (open circles) systems.  Systems
for which no Lyman limit coverage is available are shown as open
squares.  The dotted line marks the boundary between strong and weak
{\hbox{{\rm Mg}\kern 0.1em{\sc ii}}} absorbers.  Taken from Churchill
et al. (2000).}
\end{figure}
The number of strong {\hbox{{\rm Mg}\kern 0.1em{\sc ii}}} absorbers is
comparable to the number of Lyman limit systems, which implies that
nearly all strong {\hbox{{\rm Mg}\kern 0.1em{\sc ii}}} absorbers give
rise to a Lyman limit break (Churchill et al. 2000).  The number of
weak {\hbox{{\rm Mg}\kern 0.1em{\sc ii}}} absorbers (two--thirds of
which are single--cloud systems) exceeds the number of Lyman limit
systems by a factor of two.  Therefore, statistically, the vast
majority of weak {\hbox{{\rm Mg}\kern 0.1em{\sc ii}}} absorbers must
not give rise to a Lyman limit break.  This is confirmed in the right
hand panel of Figure 6, which shows that all the strong {\hbox{{\rm
Mg}\kern 0.1em{\sc ii}}} absorbers for which data were available are
found to have Lyman limit breaks.  All but one of the weak {\hbox{{\rm
Mg}\kern 0.1em{\sc ii}}} absorbers are found not to have Lyman limit
breaks.

Most all single--cloud weak {\hbox{{\rm Mg}\kern 0.1em{\sc ii}}}
absorbers are {\hbox{{\rm Ly}\kern 0.1em$\alpha$}} forest clouds in
the sense that their neutral hydrogen column densities are
$<10^{16.8}$~{\hbox{cm$^{-2}$}}.  We have discussed above that they
usually have metallicities close to solar.  By number, they comprise
between $25$ and $100$\% of the {\hbox{{\rm Ly}\kern 0.1em$\alpha$}}
forest with $15.8 < \log N({\hbox{{\rm H}\kern 0.1em{\sc i}}}) <
16.8$~{\hbox{cm$^{-2}$}} (Rigby et al. 2001).  Weak {\hbox{{\rm
Mg}\kern 0.1em{\sc ii}}} absorbers could arise in {\hbox{{\rm Ly}\kern
0.1em$\alpha$}} clouds with still smaller $N({\hbox{{\rm H}\kern
0.1em{\sc i}}})$, but this would require super--solar metallicities.
Despite the high metallicities, most of the single--cloud weak
{\hbox{{\rm Mg}\kern 0.1em{\sc ii}}} absorbers are not within $50$~kpc
of high luminosity ($>0.05L^*$) galaxies.  We conclude that a
significant portion of the {\hbox{{\rm Ly}\kern 0.1em$\alpha$}} forest
must be significantly metal enriched by redshift 0.5 to 1.0.

\section{Surprising Consequences}

We now focus on the subclass of iron--rich single--cloud weak
{\hbox{{\rm Mg}\kern 0.1em{\sc ii}}} absorbers. Consider the following
facts:

\begin{itemize}
\item{$dN/dz = 0.18$ for the subclass of iron--rich, single--cloud
weak {\hbox{{\rm Mg}\kern 0.1em{\sc ii}}} absorbers.}
\item{The inferred thicknesses of these objects are $\sim 10$~pc.}
\item{$dN/dz = 0.91$ for strong {\hbox{{\rm Mg}\kern 0.1em{\sc ii}}}
absorbers which arise in the regions within $\sim 40$~kpc of luminous
galaxies ($>0.05L^*$).}
\end{itemize}

Combining these three facts leads to the conclusion that the
structures that produce single--cloud weak {\hbox{{\rm Mg}\kern
0.1em{\sc ii}}} absorption outnumber the luminous galaxies by factor
of more than one million.  Also, recall that although that there
is no evidence they are
near luminous galaxies, the iron--rich, single--cloud weak {\hbox{{\rm
Mg}\kern 0.1em{\sc ii}}} absorbers are of solar metallicity and they
are not $\alpha$--element enhanced.  {\it In situ} enrichment by Type Ia
supernovae is implied.  It could occur within supernova remnants in
intergalactic star clusters, perhaps even the elusive Population III,
or within an abundant class of invisible dwarf galaxies.

\section{What are the single--cloud weak {\hbox{{\rm Mg}\kern 0.1em{\sc ii}}} absorbers?}

From our analysis, we infer the following:

\begin{enumerate}
\item{
The single--cloud weak {\hbox{{\rm Mg}\kern 0.1em{\sc ii}}} absorbers
are a variety of metal--rich regions in some of the same clouds that
produce {\hbox{{\rm Ly}\kern 0.1em$\alpha$}} forest lines.}
\item{
Their {\hbox{{\rm Mg}\kern 0.1em{\sc ii}}} absorption lines are weak primarily
because of a lower total hydrogen column density and a higher
ionization of the {\hbox{{\rm Mg}\kern 0.1em{\sc ii}}} phase, not
because of a lower metallicity.}
\item{
The iron--rich absorbers could be gas in star clusters or remnants
of Type Ia supernovae.  Multiphase structure (a few km/s component
and a ten km/s component) suggests that these might arise in
larger dark matter halos.}
\item{
These objects could be related to Pop III star clusters (but not
globulars because their low--redshift counterparts would easily be
detected in the Local Group) or to a population of failed dwarf
galaxies expected in CDM models.}
\item{
Weak {\hbox{{\rm Mg}\kern 0.1em{\sc ii}}} absorbers without detected
{\hbox{{\rm Fe}\kern 0.1em{\sc ii}}} may be related to Type II
supernova fragments or may be a type of intragroup HVCs.
However, as sub--Lyman limit systems they cannot be the same systems
detected as HVCs by 21--cm surveys.}
\end{enumerate}

\acknowledgements
Support for this work was provided by the NSF (AST--9617185) and by NASA
(NAG 5--6399 and HST--GO--08672.01--A), the latter from the
Space Telescope Science Institute,
which is operated by AURA, Inc., under NASA contract NAS5--26555.
N. Bond, J. Rigby, and S. Zonak were supported by an NSF REU Supplement.


\begin{references}
\reference{Bahcall, J. N., et al. 1993, \apjs, 87, 1}
\reference{Bahcall, J. N., et al. 1996, \apj, 457, 19}
\reference{Bergeron, J., \& Boiss\'e, P. 1991, \aap, 243, 344}
\reference{Charlton, J. C., Ding, J., Zonak, S. G., Bond, N. A.,
Churchill, C. W., \& Rigby, J. R. 2001, \apj, in preparation} 
\reference{Churchill, C. W., Mellon, R. R., Charlton, J. C., Jannuzi, B. T.,
Kirhakos, S., \& Steidel, C. C., 2000, \apjs, 130, 91}
\reference{Churchill, C. W., Rigby, J. R., Charlton, J. C., \& Vogt, S. S. 1999, \apjs, 120, 51}
\reference{Ferland, G. 1996, Hazy, University of Kentucky, Internal Report}
\reference{Haardt, F., \& Madau, P. 1996, \apj, 461, 20}
\reference{Jannuzi, B. T., et al. 1998, \apjs, 118, 1}
\reference{Rigby, J. R., Charlton, J. C., \& Churchill, C. W. 2001, \apj, submitted}
\reference{Steidel, C. C. 1995, in QSO Absorption Lines, ed. G. Meylan (Garching:Springer--Verlag), 139}
\reference{Steidel, C. C., Dickinson, M. \& Persson, E. 1994, \apj, 437, L75}
\reference{Steidel, C. C., \& Sargent, W. L. W. 1992, \apjs, 80, 1}
\end{references}
\end{document}